\documentclass[aps,prd,floatfix,twocolumn]{revtex4} 
\usepackage{graphicx}

\usepackage{graphicx}
\usepackage{amsmath}
\usepackage{amssymb}
\usepackage{yfonts}
\usepackage{slashed} 

\usepackage{hyperref}

\usepackage{marvosym}
\usepackage{wasysym}
\usepackage{xcolor}

\newcommand{\mmin}{{\hspace{0.2mm}-\hspace{0.2mm}}}
\newcommand{\pplus}{{\hspace{0.2mm}+\hspace{0.2mm}}}
\newcommand{\iss}{{\hspace{0.2mm}=\hspace{0.2mm}}}
\newcommand{\ew}{{\it Z}}

\renewcommand{\d}{{\rm d}}

\newcommand{\cm}{{\rm cm}}

\newcommand{\diag}{{\rm diag}}

\newcommand{\half}{{\frac{1}{2}}}

\newcommand{\myskip}[1]{}

\newcommand{\tr}{{\rm tr}\,}

\newcommand{\onedot}{\,\,\,}   

\newcommand{\ed}{\onedot}%{\,\,\cdot\,}   
\newcommand{\ednu}{{\onedot\nu}}%{\,\,\cdot\,}   

\newcommand{\BEQ}{\begin{eqnarray}}   
\newcommand{\EEQ}{\end{eqnarray}}   
\newcommand{\BEA}{\begin{eqnarray}}   
\newcommand{\EEA}{\end{eqnarray}}   
\newcommand{\nn}{\nonumber }   
\renewcommand{\d}{{\rm d}}

\renewcommand{\th}{\vartheta}   
   
\newcommand{\eps}{\varepsilon}

\newcommand{\gam}{\gamma}   

\newcommand{\mn}{{\mu\nu}}

\renewcommand{\diag}{{\rm diag}}

\newcommand{\cA}{{\cal A}}
\newcommand{\cB}{{\cal B}}
\newcommand{\cC}{{\cal C}}

\newcommand{\cK}{{\cal K}}
\newcommand{\cL}{{\cal L}}

\newcommand{\sth}{s_\theta}

\newcommand{\cW}{c_W}
\newcommand{\sW}{s_W}

\newcommand{\ih} {{ \it i}}

\begin{document}
\draft

\title{The interior of hairy black holes in standard model physics}

\author{Theo M. Nieuwenhuizen}

\affiliation{Institute for Theoretical Physics, 
University of Amsterdam \\
% Science Park 904,  
PO Box 94485,  1090 GL Amsterdam, The Netherlands}

% \maketitle 

\date{%Today=\today; 
Version: August 3, 2021}
%\shortauthor{Th.M. Nieuwenhuizen}

%{Fundamental problems and general formalism}
%{Canonical formalism, Lagrangians, and variational principles} 
%{Origin and formation of the Universe}

\begin{abstract}
A large class of stationary, non-rotating  black hole metrics is proposed, in which the interior is regular with a core 
consisting of a condensate of Higgs and $Z$ bosons generated from the nuclear binding energy of the initial H atoms.
Gravitational collapse is prevented by negative pressures from the Higgs condensate and a small imbalance 
in the distribution of electric charges.
Non-condensed, thermal particles are present as well. The  approach holds for masses exceeding  $0.75\,10^{-4}M_\odot$.
The inner horizon sets an inner core of 11 cm, while the characteristic radius of the full core is  $270\,(M/M_\odot)^{1/3}$ cm.
For increasing charge, the core expands; in the extremal case, it fills the interior. 
While the net charge is easily shielded, the build up of horizons may prevent this in the interior, 
and consequently avoid a singularity.
In black hole merging, the core of a nearly extremal one may be exposed, forming a new class of events.
The approach is a stepping stone towards rotating black holes.
\end{abstract}

\maketitle

 \section{Introduction}

A black hole (BH) is said to have  an event horizon but no hair\cite{misner1973gravitation},
being described only by its mass, charge and angular momentum. 
Nothing can escape from it, except for Hawking radiation\cite{hawking1974black}, a feeble quantum effect.
The 2020 Nobel prize  was awarded to Roger Penrose ``for the discovery that black hole formation is a robust 
prediction of the general theory of relativity", 
and to Reinhard Genzel and Andrea Ghez ``for the discovery of a supermassive compact object at the centre of our galaxy''. 

A BH is a trapped surface, often described by a metric. 
The Schwarzschild metric\cite{schwarzschild1916gravitationsfeld}  involves only the mass; the Reissner--Nordstr\"om  (RN) metric\cite{reissner1916eigengravitation,weyl1917gravitationstheorie,nordstrom1918energy,jeffery1921field}  adds a central charge,
and the Kerr--Newman (KN) metric\cite{newman1965note}  models its rotation. When charge free, the rotating BH is described by the
Kerr metric\cite{kerr1963gravitational}, widely employed  in astrophysics.
These metrics are source free, that is, the matter density vanishes everywhere, except at the singularity.
The Schwarzschild and Reissner-Nordstr\"om metrics are singular at the origin.
It is widely believed that the singularity must be described by quantum gravity\cite{nobel2020,rovelliEPnews2021}.

 Regularized metrics have been proposed, see, e.g., \cite{bardeen1968non,hayward2006formation,nieuwenhuizen2008exact,nieuwenhuizen2010bose, frolov2014information,casadio2015thermal}, 
 and the reviews \cite{mazur2015surface,simpson2020regular}. 
 They were presented in an ad hoc manner, but the Bardeen black hole 
(and its generalizations) can be built self-consistently by considering a coupling to non-linear electrodynamics \cite{ayon2000bardeen}. 

While these attempts involve hypothetical forms of matter, we propose, as next step forward,  
a large class of regular BH metrics involving the standard model (SM) of particle physics.
The SM is our best theory of Nature; it has been vindicated for dozens of reaction channels at the LHC\cite{CMSPreliminary}.
Employing the SM  for BH interiors would vindicate it the more and save us from having to open
``Pandora's box'' of paradoxes related with the BH singularity\cite{carballo2020opening}. 
 Last but not least, it would even do away with the need for quantum gravity in astrophysical BHs.

 In the present work we consider models for charged, non-rotating BHs.
In section II we lay out the basic approach, while we treat the full approach in section III. 
In section IV we give a summary and in section V we discuss the setup and speculate on possible merits.

 \section{The basic approach}

The Schwarzschild metric describes a point mass in GR (general relativity).
With $S(r)=2GM/r$, \! it reads\footnote{\label{Unist} %In our %We employ units,
We set $c=\hbar=1$,  $\mu_0=4\pi$, so that $e^2=\alpha=1/137$.
The Planck mass is $m_P=1/\sqrt{G}$. $m_N$ is the nucleon mass. \label{units}} 
 \BEQ\label{gSchw}
 g_\mn={\rm diag}(1-S,\frac{1}{S-1},-r^2,-r^2\sth^2),\quad \sth=\sin\theta ,
 \EEQ
where $r^\mu=(t,r,\theta,\phi$) with $\mu=0,1,2,3$ denote  areal-like coordinates.
Signals get an infinite redshift at the event horizon $r=R_S=2GM$, where $g_{00}\to 0$;
this astonished Einstein so much, that he initially dismissed the case.

The RN BH with charge $Q$ is covered by $M\to \mu (r) =M-Q^2/2r$.
The Kretschmann curvature invariant,
\BEQ && \hspace{-4mm}
\cK=R^{\kappa\lambda\mu\nu}R_{\kappa\lambda\mu\nu}=(4G^2/r^6)\times \\
&&  \hspace{-6mm}
\big(12 \mu  ^2-16 r\mu  \mu  '+r^2 \mu  \mu  ''+8 r^2\mu  '{}^2-4 r^3 \mu  ' \mu  ''+r^4\mu  ''{}^2\big),  \nn
\EEQ
diverges for $\mu(r)  =M$ as $1/r^6$ for $r\to0$, and in the RN case even as $1/r^{8}$, 
which points at a physical singularity at the origin.
This has led to the opinion that the BH singularity can only be described by  quantum gravity.

Since $\cK$ is regular at $R_S$,  no physical singularity occurs there.
Indeed, an observer freely falling into the BH
is described by the Painlev\'e Gullstrand (PG) metric\cite{painleve1921mecanique,gullstrand1922allgemeine},  which is regular for $r>0$, viz.
$g_\mn=\gam_\mn -S k_\mu k_\nu$,  where $\gam_\mn=\diag(1,-1,-r^2,-r^2\sth^2)$  is the Minkowski metric and
$k_\mu=(1,1,0,0)$ a null vector, viz. $g^\mn k_\mu k_\nu=0$.

To study the BH interior, we consider a general shape of $\mu(r)$. The Einstein equations then do not deal with a vacuum situation 
but demand an energy momentum tensor (EMT)
\BEQ \label{TMn=}\label{cAcB}
&&
T^\mu_{\ed\nu}
=\cA \,\delta^\mu_{\ed\nu}+\cB \, \cC^\mu_{\ed\nu} ,\quad  \cC^\mu_{\ed\nu} =\diag(1,1,-1,-1)  , \nn\\&&
\quad \cA=\frac{2\mu  '+r\mu  ''}{16\pi r^2} ,\qquad 
\quad \cB=\frac{2\mu  '-r\mu  ''}{16\pi r^2}  .
\EEQ
 This result holds for both metrics, since they differ by their time variable while both are stationary.

\subsection{The mass in the field theoretic approach}

The {\it mass} of the metric is often derived by invoking the far field only. 
In the field theoretic description of gravitation, a bulk description is achieved.
Hereto one assumes an underlying Minkowski space with metric $\gam_\mn$ and
one introduces $k^\mn=\sqrt{g/\gam}\,g^\mn$ with $g=\det g_\mn$ and $\gam=\det\gam_\mn$. 
The Einstein equations can then  be formulated in the Minkowski space \cite{landau1975classical,babak1999energy}
\BEQ\label{Amn=}
A^\mn=8\pi G\Theta^\mn,\quad A^\mn=\frac{1}{2}(k^{\mu\nu}k^{\alpha\beta}-k^{\mu\nu}k^{\alpha\beta})_{:\alpha\beta} ,
\EEQ
where the ``acceleration tensor'' \cite{nieuwenhuizen2007einstein} $A^\mn$ involves covariant derivatives in Minkowski space, 
indicated by the column. The total energy momentum tensor is $ \Theta^\mn=(g/\gam)(T^\mn+t^\mn )$, 
with $t^\mn$ representing the EMT of the gravitational field; in Cartesian coordinates it is the Landau-Lifshitz pseudo tensor.
In our case:  $k^\mn=g^\mn$, since $g=\gam=-r^4\sth^2$.  The mass (energy) of the metric is
\BEQ \label{Eintegral}
E=\int_{R^3}\d^3r\sqrt{-\gam}\Theta^{00},\quad \d^3r=\d r\d \theta\d \phi .
\EEQ 
For the PG metric this results in $E=\mu  (r)\big|^\infty_0 $.
The general approach to get $E=M$ is to take the value at $r\to\infty$ and neglect the value at  $r=0$.
For the metric (\ref{gSchw}) the equivalent is $E=\mu(r)/(1-2G\mu(r)/r)\big|^\infty_0$, 
which, instead, exposes a peculiarity at $R_S$.
So a fundamental approach has to deal with the interior, in particular with $r\to 0$.

\subsection{Employing the standard model} 

We shall investigate whether the standard model {\it can} describe the BH interior.
Clearly, $\cK$ remains finite when $\mu  \sim r^3$ for $r\to 0$, in which case $E=M$, as desired.
In the dynamical BH formation process, this is the worst case before a singularity develops.
From (\ref{TMn=}) it follows that $\int _{R^3}\d^3 r\,T^0_{\ed 0}=\int_0^{R_S} \d r\,4\pi r^2(\cA+\cB)$ indeed equals 
$\mu(R_S)=M$, since $\mu(0)=0$ grace to the regularization.

As in the RN case, the condition $S=1$ implies an event horizon (at $R_S$) and an inner horizon at some $R_i>0$.

The $\cB$ term in (\ref{TMn=}) may arise from an electric charge distribution. Consider a potential $A_\mu=A_0(r)\delta^0_\mu$. 
It derives from a charge density [35]
\BEQ \label{Jmu=}
J_\gam^\mu=J_\gam^0\delta^\mu_{0},\quad J_\gam^0=-\frac{1}{\mu_0}(A_0''+\frac{2}{r}A_0') .
\EEQ
It has charge $Q=-R_S^2A_0'(R_S)$ and Coulombian EMT, 
\BEQ
T^\mu_{\gamma\nu}=\frac{A_0'{}^2}{8\pi}  \cC^\mu_{\ed\nu}  , \qquad  \cC^\mu_{\ed\nu} =\diag(1,1,-1,-1)  .
\EEQ
For the RN metric: $A_0=Q/r$, $\cA=0$ and $\cB=Q^2/8\pi r^4$.
In general, matching with (\ref{TMn=}) imposes
\BEQ\label{A0p=}
 -A_0'  =\sqrt{\frac{2\mu  '-r\mu  ''}{2r^2} }  =\frac{m_P}{2r} \sqrt{2S-r^2S''}, \hspace{1mm} S\equiv\frac{2G\mu}{r},
\EEQ
with $m_P=1/\sqrt{G}$ the Planck mass.
The small-$r$ refinement $\mu  \sim r^3-r^5$ implies regularity, $A_0'\sim- r$ and $\cB\sim r^2$.
The charge is zero when $r\mu''=2\mu'$ ($=0$ typically) at $R_S$.

Also the $\cA$ term  in (\ref{TMn=}) may result from SM physics.
Inside the BH, the matter, mostly H-atoms, is dissolved into electrons and quarks, 
with the nuclear binding energy converted, in principe, into Higgs, $Z$ and $W$ bosons, photons and gluons.
To start, we assume that the interior involves a condensate of the Higgs field\footnote{The normal vacuum is a Higgs condensate, 
compensated by a vacuum term; in the BH the compensation is lost.}.
At  the macroscopic scale, {\it the kinetic terms of the massive fields can be neglected},
as they bring a factor $1/R_S^2$ $=m_P^4/M^2$ rather than $M_{H,Z,W}^2\sim v^2$. 
In the absence of  $W$ and $Z$ fields,  the EW Lagrangians reads, in the unitary gauge,
\BEQ \hspace{-1mm}
\cL_H=-\frac{\lambda}{4}(\varphi^2-v^2)^2-\frac{\varphi}{v}\rho_F ,\hspace{4mm}
\cL_\gamma=\frac{F^\mn F_{\nu\mu}}{4\mu_0},
\EEQ
with $v=246$ GeV the vacuum expectation value of $\varphi=v+h$, with $h$ the Higgs field, and  $\lambda=0.129$ the self-coupling.
Next, $\rho_F$ is the rest mass density of the fermions (electrons, and up and down quarks), with their Yukawa coupling to $\varphi$ made explicit.
Neglecting the variation of $J^\mu_\gam$ with the metric,
the EMTs have elements [14] 
\BEQ \label{TewMn=} \hspace{-2mm}
T^\mu_{H\nu}=- \delta^\mu_\nu  \, \cL_H ,\hspace{2mm}
T^\mu_{\gamma\nu}= \frac{1}{\mu_0}F^{\mu\rho}F_{\rho\nu}-\delta^\mu_\nu \, \cL_\gam.
\EEQ
It yields $\cA_H =\frac{1}{4}\tr T_H=\frac{\lambda}{4}(\varphi^2-v^2)^2+\frac{\varphi}{v}\rho_F$.
Upon matching it with $\cA$, the Higgs field gets pushed away from its vacuum value $h=0$.
Since its derivatives are negligible, $\varphi$ is subject to the equation of motion
\BEQ\label{phimin}
\lambda(\varphi^2-v^2)\varphi=-\frac{\rho_F}{v} ,
\EEQ
 the equivalent of the Gross-Pitaevskii equation\cite{pitaevskii2016bose} 
 for the condensate wavefunction in Bose-Einstein condensation.

\subsection{A specific example}

The positive energy conditions $\cA,\cB\ge0$, with the conditions at $0$ and $\infty$, constrain the unspecified function $\mu$.  
Nevertheless, realistic cases exist; an example is
\newcommand{\rs}{r_s}
\BEQ \hspace{-2mm} \label{muexample}
\mu  = \frac{M_1 r^3 }{(\sqrt{r_s^2 + r^2} + r_u)^3} ,\hspace{1mm}
M_1=M \frac{(\sqrt{r_s^2 + R_S^2} + r_u)^3}{R_S^3} , \hspace{1mm}
\EEQ
where $r_u=ur_s$ with $u$ a $c$-number. This yields
\BEQ
\cA&=&\frac{3 M_1x^5}{16 \pi  \rs^3 u^7}\frac{ 4 u^4-u^2+7 u^2 x+5 x^2+x^3}{ (x+1)^5}, \qquad \\
\cB&=&\frac{3 M_1x^6 }{16 \pi  \rs^3 u^7}\frac{r^2}{\rs^2}\frac{ 4 u^2+5 x+x^2}{ (x+1)^5} , \quad
 x=\frac{r_u}{\sqrt{\rs^2+r^2}}  . \nn
\EEQ
While $\cA(0)$ and $\cB(r)$ are positive, the positivity of $\cA(R_S)$ is secure when $u\ge \half$.
  The Kretschmannian,
\BEQ 
&& \hspace{-7mm} 
\cK=\frac{24 G^2M_1^2 x^6 }{\rs^6 u^{14} (x+1)^{10}} \Big[ \, 4 u^8 (2 x^4+7 x^2-2 x+1) +
\nn \\ && \hspace{-7mm} 
4 u^6 x^2 (7 x^3-11 x^2+23 x-7)+6 u^2 x^6 (3 x^2+10 x-25)
\nn \\ && \hspace{-7mm} 
+\,u^4 x^4 (4 x^3+43 x^2-142 x+107)+3 x^8 (x+5)^2\Big], 
\EEQ
is finite and maximal at $r=0$. The EM  potential reads
\newcommand{\bv}{\bar u}
\BEQ
 A_0&=&\textrm{const.}-\frac{4}{3}\sqrt{\frac{ M_1}{\rs u^3}}\Big[ \,
\frac{\sqrt{6\,(4u^2+5x+x^2)}}{4(1+x)^{3/2}} \\
&+& \, \frac{E\big(i\sqrt{\frac{3(1+\bv)}{2(1+x)}};\sqrt{\frac{1-\bv}{1+\bv}}\big)
-\bv\,F\big(i  \sqrt{\frac{3(1+\bv)}{2(1+x)}};\sqrt{\frac{1-\bv}{1+\bv}}\big)}{ i (1-\bv)\sqrt{1+\bv}}
\Big] , \nn
\EEQ 
with the const. such that $A_0(0)\equiv0$ and with the incomplete elliptic integrals of the first and second kind,
\BEQ
E(iw;k)-\bv F(iw;k)=i\int_0^w\frac{\d t\,(1+k^2t^2-\bv)}{\sqrt{(1+k^2t^2)(1+t^2)}} ,
\EEQ
and $\bv= \sqrt{25 - 16 u^2}/3$. The charge density,
\BEQ
&&  \hspace{-5mm}J^0=\frac{\sqrt{6 M_1} }{16 \pi  u^{5/2} }\,x^4 \times \\&& \hspace{-5mm}
\frac{5u^2(4u^2-1)+u^2 (7 x^2+42 x)+x^2 (3 x^2+18 x+35)}{ u^3(x+1)^{7/2} \sqrt{4 u^2+5x+x^2}}  , \nn
\EEQ
 is nonnegative. The charge within radius $r$ is
 \BEQ
 Q(r)= 
 \sqrt{6M_1r_u } \, \frac{(u^2-x^2)^{3/2} \sqrt{4 u^2+5x+x^2}}{2u^4 (x+1)^{5/2}} ,
 \EEQ 
 and the total BH charge is $ Q=Q(R_S)$.

\subsection{Physical estimates}

Equating  the average mass density $M/R_S^3$ to the EW one, $\lambda v^4$, defines, surprisingly, 
a Neptunian mass scale
\BEQ
M_\ast=\frac{m_P^3}{\sqrt{\lambda}v^2}=0.75\times 10^{-4}M_\odot=25M_\oplus=1.5M_\text{\Neptune} .
\EEQ
For $M\gg M_\ast$, let $\mu  (r)$ go from 0 to $M$ at a typical core scale $R_c\ll R_S$.
An EW mass density up to $R_c$ then sets 
\BEQ\hspace{-3mm}
R_c\sim \frac{M^{1/3}}{\lambda^{1/3}v^{4/3}} = R_\ast \,\big(\frac{M}{M_\ast}\big)^{1/3}
=260 \,\big(\frac{M}{M_\odot}\big)^{1/3} \, \cm  .
\EEQ
With $\mu \sim M (r/R_c)^3$, the inner horizon $R_i=2G\mu(R_i)$ 
corresponds to a fixed characteristic scale $R_\ih\approx R_\ast$,  viz.
\BEQ
R_\ast=GM_\ast=\frac{m_P}{\sqrt{ \lambda} v^2}=11\,\cm.
\EEQ
The condition $R_c\gg R_\ast$ again implies $M\gg M_\ast$.

Like for the Earth, we call the sphere $r<R_\ih$ {\it the inner core}, 
the spherical shell $R_\ih<r<R_c$ {\it the outer core} and the region $R_c<r<R_S$  {\it the mantle}. 
The transition from inner to outer core is sharp,  from outer core to mantle 
it is gradual. Instead of a crust, there is an event horizon. 
 
The dimensionless value of $\cK$ in this smeared singularity shows that curvature is strong but finite,
\BEQ
R_c^4\cK\sim \frac{R_S^2}{R_c^2}\sim \big(\frac{M}{M_\ast}\big)^{4/3} \sim 3\times10^5\,\big(\frac{M}{M_\odot}\big)^{4/3} .
\EEQ

The electromagnetic field has energy density $A_0'{}^2\sim \lambda v^4 $.
The typical values of $A_0$ and the total charge $Q$ are
\BEQ \hspace{-3mm}
 |A_0|\sim \sqrt{\lambda}v^2R_c =(\frac{M}{M_\ast})^{1/3}m_P,\hspace{2mm}
 Q\sim (\frac{M}{M_\ast})^{2/3}\frac{m_P^2}{\sqrt{\lambda}v^2}.
\EEQ
 A RN BH has maximal charge $Q=M/m_P$.
The fraction of uncompensated charges that form the BH,
 $\eps=(Q/e)/(M/m_N)$, has typically the small value 
 $\eps_M=(m_N/em_P)(M_\ast/M)^{1/3}= 4\,10^{-20} (M_\odot/M)^{1/3}$,
which is still large enough to prevent collapse into a singularity.
 
Lastly, the fraction of rest mass in the fermions is $(2m_u+m_d+m_e)/m_N=0.0079$.
This implies that the $\cA_H$ below eq. (\ref{TewMn=}) can not match $\cA$ numerically.

\section{The full SM approach}

Having just run into a numerical mismatch, we have  to  widen our search for  an EW description of the BH interior. 
We do this in 3 ways. 1) Allow also condensed $Z$ bosons ($W$ bosons perform less well); 
2) Allow for thermal particles:
thermal electrons and quarks, Higgs and $Z$ bosons, and photons and gluons. 
3) Dilatate time by setting $\d t\to S_0\d t$, $g_{00}\to (1-S)S_0^2$ with $S_0(r)\le1$.
Since the BH is stationary,  its mass will  not depend on $S_0$.
  
1)  {\it The $Z$ condensate.}
We extend the EW Lagrangian and EMT by the $Z$ field contribution\BEQ \hspace{-3mm}
\cL_Z \iss \frac{M_Z^2\varphi^2}{2\mu_0v^2}Z^\rho Z_\rho, \hspace{1mm}
T^\mu_{Z\,\nu}\iss \frac{M_Z^2\varphi^2}{\mu_0v^2}\big(Z^\mu Z_\nu \mmin \dfrac{\delta^\mu_\nu}{2} Z^\rho Z_\rho\big).
\EEQ

2) {\it Thermal matter.} 
At finite temperature, a Bose-Einstein condensate does not collect all particles, leaving {\it thermal} particles in the
 gas\cite{pitaevskii2016bose}.
A thermal EMT has the general form $T^\mu_{\th\,\nu}=(\rho_\th+p_\th)U^\mu U_\nu-p_\th\delta^\mu_\ednu$ with $U^\mu$ the
velocity vector. In the inner core, time is normal so that $U^\mu=\delta^\mu_{\ed 0}/\sqrt{g_{00}}$ 
and $T^\mu_{\th\,\nu}=\diag(\rho_\th,-p_\th,-p_\th,-p_\th)$.
In the outer core and mantle, $r$ is the time-like variable, so that $T^\mu_{\th\,\nu}$ $=$ $\diag(-p_\th,\rho_\th,-p_\th,-p_\th)$.
This combines as 
\BEQ
&& \hspace{-5mm} 
T^\mu_{\th\,\nu}=\cA_\th \delta^\mu_\ednu+
\frac{\sigma}{4\pi G}\,\cC^\mu_\ednu+\frac{\tau}{2\pi G}(\delta^\mu_0\delta^0_\nu-\delta^\mu_1\delta^1_\nu)
,  \\
&& \hspace{-5mm}
 \cA_\th = \frac{\rho_\th- 3p_\th }{4}, \quad 
 \sigma=\pi G(\rho_\th+p_\th), \quad \tau=\pm\,\sigma, \nn
\EEQ
with $\tau=\sigma$ in the inner core but $-\sigma$ in the outer core and mantle.
Relativistic thermal particles lead to $\cA_\th=0$.

3) {\it Time dilatation}. For $S_0(r)\neq 1$, the trace of  $T$ brings 
\BEQ
 \cA=\frac{2S+4rS'+r^2S''}{32\pi Gr^2} +
\frac{3S'_0S'}{32\pi GS_0 } 
+ \frac{2S_0'+rS_0''}{16\pi G rS_0}\bar S ,
\EEQ 
where $\bar S=S-1$. The enclosed charge is $Q=-r^2A_0'/S_0$. 

The equation of motion for the Higgs field (\ref{phimin}) becomes
\BEQ\label{phimin1}
\lambda(\varphi^2-v^2)\varphi=\frac{\varphi}{\mu_0v^2}M_Z^2Z^\rho Z_\rho-\frac{\rho_F}{v} .
\EEQ
Having pointed out that $\rho_F$ is small, we now neglect it, so that $\varphi$ is a square root,
and determines
\BEQ
\cA_\ew \equiv \frac{1}{4}{\rm tr}\, T_\ew =-\frac{1}{4\mu_0}\frac{\varphi^2}{v^2}M_Z^2Z^\rho Z_\rho.
\EEQ
We assume that $Z_\mu=\delta^0_\mu Z_0$, whence $\cA_\ew$ is negative in the inner core, and positive in outer core and the mantle.
Equating $\cA$ to $\cA_H+\cA_\ew+\cA_\th$ yields solutions when $\cA_\th>0$;
this is possible when the temperature $T$ is not too high, so that the Higgs and $Z$ bosons are not fully relativistic.

Matching the traceless parts of $T$ and $T_\gam+T_\ew+T_\th$ yields
\BEQ \label{A0psq}
&& \hspace{-4mm}
\frac{A_0'{}^2}{S_0^2}=\frac{2S \mmin r^2S''}{4Gr^2} \mmin \frac{ 3S_0'S'}{4GS_0} \pplus \!\frac{S_0'\mmin rS_0''}{2GrS_0}\bar S
-\frac{\sigma-\tau}{2G} ,
\\&&  \hspace{-4mm}
\label{Z0sq}
\frac{M_Z^2\varphi^2}{m_P^2v^2}\frac{Z_0^2}{S_0^2}
=\bar S^2\frac{S_0'}{rS_0}+ \bar S\tau . \hspace{-1cm}
\EEQ

 Since $\bar S\tau\le0$ both in the core and the mantle, thermal matter is allowed when $S_0'>0$,
 increasing from $S_0(0)<1$ to $S_0(R_S)=1$. The small-$r$ expansions $S=s_2r^2-s_4r^4$,
 $S_0=u_0(1+u_2r^2+u_4r^4)$ and ${\color{black}\sigma=\tau=\sigma_2r^2}$ yield
 \BEQ
&& \cA=3\frac{ s_2-u_2}{8 \pi G}+\frac{12 s_2 u_2-15 s_4+6 u_2^2-20 u_4}{16 \pi G}r^2 , \nn\\
&& \frac{A_0'{}^2}{u_0^2}=\half\left(5s_4 +8 u_4-6 s_2 u_2\right)r^2,
 \\ &&
\frac{M_Z^2\varphi^2}{m_P^2v^2}\frac{Z_0^2}{u_0^2}=2u_2-(4 s_2 u_2-2u_2^2-4 u_4+\sigma_2)r^2  , \nn
 \EEQ
which exhibits a large class of proper, positive solutions with nontrivial $S_0$, viz.  $0<u_2<s_2$ or $u_2=0$, $u_4>\frac{1}{4}\sigma_2$.

At finite $r$, one example is $S=2G\mu/r$ from  (\ref{muexample}) and  $S_0=(r^2+u_0R_1^2)/(r^2+R_1^2)$.
It is natural to put $\sigma=|\bar S|\bar \sigma$,  $\tau=-\bar S\bar \sigma$ for some $0\le \bar\sigma(r)\le S_0'/rS_0$, which sets $Z_0$ via (\ref{Z0sq}).
Hence $\tau$ and $\sigma$ vanish at $R_i$, and  (\ref{A0psq}) is continuous.

\subsection{$Z$ and $W$ sources}   \label{ZWsources}

While the above solution allows that $Z_0=0$, the general case 
demands that it is properly sourced. Its equation of motion $M_Z^2(\varphi^2/v^2)Z^\mu=-\mu_0J_Z^\mu$ implies
a number of unbalanced particles 
$\delta N_Z\sim R_c^3J_Z^0\sim (M/M_\ast)^{2/3} M_Zm_P^3/\lambda v^4=2.0\times 10^{53}(M/M_\odot)^{2/3} $,
well less than the $N_H\sim 1.2\times 10^{57}(M/M_\odot)$ H atoms that make up the BH.
 For a $W$ boson field $W_\mu=W^\dagger_\mu$ instead of $Z_\mu$, 
the same would apply; however, this is may not extend to the rotating case.
Then the quartic terms of the EW Lagrangian contain a term $A^3A_3 W^ 0W_0  \sim  A_0^2 W_0^2$ 
which overwhelms the $M_W^2W_0^2$ term. As a result, the number of $W$-sourcing particles 
$\delta N_W\lesssim (M_Z/A_0) \delta N_Z\sim 10^{-17}(M_\ast/M)^{1/3}\delta N_Z$ is negligible,
while their contribution to the EMT is quadratically small in $v/m_P$.

The fermion-boson coupling    $\cL=A_\mu J^\mu_\gam+Z^0_\mu J_Z^\mu$ 
involves the electrodynamic and neutral currents (see, e.g., Peskin and Schroeder, eq. (20.80) \cite{peskin2018introduction}),
\BEQ && \hspace{-4mm}
J^\mu_\gam =\frac{2e}{3}(n^\mu_{u_L}+n^\mu_{u_R})-\frac{e}{3}(n^\mu_{d_L}+n^\mu_{d_R})-e(n^\mu_{e_L}+n^\mu_{e_R}),
\nn\\&& \hspace{-4mm}
J^\mu_Z=\frac{g}{2\cW}\big(n^\mu_{u_L}-n^\mu_{d_L}-n^\mu_{e_L}+n^\mu_{\nu_{eL}}\big) 
- \frac{\sW}{\cW}J^\mu_\gam .
% - \frac{g\sW^2}{e\,\cW}J^\mu_\gam .
\EEQ
where $n^\mu_x\equiv \bar x\gam^\mu x$ for up and down quarks, electrons and neutrinos,
with left ($L$) or right ($R$) handed chirality. 
Here $e$ is the proton charge, $g=e/\sW$ the weak coupling constant,
$\sW=\sin\theta_W$ and $\cW=\cos\theta_W$.

These currents vanish for a star consisting only of ionized H atoms, 
$n_{u_h}^\mu=2n^\mu_{d_h}=2n_{e_h}^\mu$ for $h=L,R$, while $n^\mu_{\nu_{eL}}=0$.
 In a supernova explosion more electrons get expelled  than protons, which makes $J_\gam$ finite; likewise, expulsion 
of neutrinos (e.g. from the proton-proton reaction $p+p\to d+e^++\nu_e$) makes also the first term of $J_Z$ finite.

 \section{Summary}

To offer an alternative to the present unresolved structure of the black hole (BH) singularity,
we set out  to describe the interior of a BH by extended Bose-Einstein condensates in standard model physics.
While it is standard to go to a Tolman-Oppenheimer-Volkov equation,
in our situation it is  more adequate to work with the Einstein equations themselves, where all matter components
can be specified separately and we actually made some progress for the rotating case, to be discussed elsewhere.

To start, we recall that the  Painlev\'e Gullstrand metric, which is regular at the event horizon, 
leads to an ill defined value of the mass and divergent curvature invariant.
Replacing the mass $M$ by some smooth function $\mu(r)$ with $\mu\sim r^3$ for $r\to 0$,  heals this, 
at the cost of a nontrivial energy momentum tensor.  
We seek to build the latter  from the electroweak sector of the standard model of particle physics.
This puts forward that the BH interior contains a Higgs and $Z$ condensate with mass densities comparable 
to the electroweak one, $\lambda v^4$.

Gravitational collapse is prevented by negative pressures,
an isotropic one from the Higgs and $Z$ condensates and a radial one from  the Coulomb repulsion
due to a small mismatch of electric charges.

In the thus considered picture, the H atoms that form the BH, dissolve into electrons, and up and down quarks. 
 Elementary reactions like $u+e\leftrightarrow u+ W_-+\nu_e\leftrightarrow d+\nu_e$  take place, next to
 the creation of Higgs and  $Z$ bosons, and photons and gluons, from the nuclear binding energy.
 
In the simplest approach, the core of the BH involves 
 a Higgs condensate and an electrostatic potential, but it appears that the source for the condensate is too small.
This is overcome by also allowing condensed $Z$ bosons and various thermal particles, as well as a local time dilatation coded in the function $S_0(r)$.

 A nontrivial value for the $\cA_\th=(\rho_\th-3p_\th)/4$ of the thermal particles acts as a positive energy condition;
 photons, gluons and relativistic quarks and electrons do not contribute to this, hence
it requires that the temperature be $\lesssim\lambda^{1/4}v$ to prevent
that also the thermal Higgs and $Z$ particles are relativistic and $\cA_\theta$ too small.

 \section{Discussion and outlook} 
 
The Planck mass and the microscopic $v$ set a characteristic, Neptunian mass scale,
$M_\ast=m_P^3/\sqrt{\lambda}v^2=0.75\times 10^{-4}M_\odot=1.5M_\text{\Neptune}$.
Our approach holds for larger masses, which is on the safe side given that  BHs of stellar origin are neither expected  nor observed
with mass less than solar. The related characteristic length scale,  $R_\ast=GM_\ast=m_P/\sqrt{\lambda}v^2=11$ cm,  
sets the size of the inner core, the region inside the inner horizon, where time has its normal meaning; in the outer core and in the mantle,
the radial coordinate has the meaning of time, so that matter can get only ever closer to the origin. 

The  mass is located in the combined inner and outer core; it has characteristic size $R_c=(M/M_\ast)^{1/3} R_\ast$.
Beyond it, up to the much larger Schwarzschild radius, there is essentially the vacuum envisioned by Schwarzschild.
Our solution provides consistent particle distributions given the $\mu$ and $S_{0}$ profiles. 
A criterion for optimizing the profiles is desired.

So far, our charged BH model can act as stepping stone for the case of rotating BHs.
But let us consider its own merits in a speculative way.
 In a supernova explosion, electrons are more easily ejected than protons, making the matter for the BH positively charged.
In an imploding stellar core of mass $M$ and charge $Q$, with profiles $M(r)$ and $Q(r)$ and uniform charge--to--mass ratio,
the ratio of forces on a chunk ($ch$) of matter, $[Q(r)Q_{\it ch}/r^2]/[GM(r) M_{\it ch}/r^2]=(Qm_P/M)^2$, 
allows BH formation just up to the extremal charge $M/m_P$. 
For our value $(Qm_P/M)^2\sim0.002\,(M_\odot/M)^{2/3}$, this back-of-the-envelope estimate 
offers wiggle room for the onset of charged, few solar mass BH formation.
While the net charge is easily undone by pair creation near the horizon or by accumulation of negative charges from the
accretion disk, see e.g., \cite{ghosh2021astrophysical}, this may not be achieved locally during the collapse,
and subsequently made impossible by the establishment of horizons,
in which case collapse into a singularity can be avoided. The locked-up internal net positive charge
gets compensated by a negatively charged layer just outside the event horizon.
If accretion makes the BH extremal  (in charge, combined with rotation),
the horizons disappear and fireworks occur.

Typically, the mantles are much larger than the cores, so that the cores are shielded  in BH-BH merging.
If one of the BHs is nearly extremal,  
ejection of matter may occur due to the exposition of its large core.
This predicts a new class of events, alternative to neutron star-BH mergers,  with a mass allowed 
above the neutron star maximum of $\sim2M_\odot$;
a candidate for this is S190426c\cite{Ligo2019a,Ligo2019b,lattimer2019properties}.

Another possibility is that the  ejecta get a high energy.
Indeed, the Coulomb energy of an isolated charge $+\,e$ at the extremal event horizon radius $R_g=GM$ equals
$Qe/R_g=Qm_P^2\sqrt{\alpha}/M\approx 10^{27}(Qm_P/M)$ eV. When this is not screened strongly and cascades,
such BH-BH mergers offer an explanation for ultrahigh energy cosmic rays and neutrinos,
 coincident with the emission of radiation and gravitational waves.

The present work offers a new view on astrophysical BHs and a further application of the standard model of particle physics.
Many questions remain, like:  Can this class of metrics be reached dynamically?
Are they (meta)stable?  What is their entropy? Can the approach be generalized to rotating BHs?
Are there connections with the Fermi bubbles and/or the X-ray chimneys around Sag A$^\ast$?
\myskip{Are there observable effects of expelled charges?}%%%

\acknowledgements

Discussion with Gregory Gabadadze and correspondence with Sera Markoff, as well as remarks by 
 anonymous referees, are gratefully appreciated.


\begin{thebibliography}{33}%
\makeatletter
\providecommand \@ifxundefined [1]{%
 \@ifx{#1\undefined}
}%
\providecommand \@ifnum [1]{%
 \ifnum #1\expandafter \@firstoftwo
 \else \expandafter \@secondoftwo
 \fi
}%
\providecommand \@ifx [1]{%
 \ifx #1\expandafter \@firstoftwo
 \else \expandafter \@secondoftwo
 \fi
}%
\providecommand \natexlab [1]{#1}%
\providecommand \enquote  [1]{``#1''}%
\providecommand \bibnamefont  [1]{#1}%
\providecommand \bibfnamefont [1]{#1}%
\providecommand \citenamefont [1]{#1}%
\providecommand \href@noop [0]{\@secondoftwo}%
\providecommand \href [0]{\begingroup \@sanitize@url \@href}%
\providecommand \@href[1]{\@@startlink{#1}\@@href}%
\providecommand \@@href[1]{\endgroup#1\@@endlink}%
\providecommand \@sanitize@url [0]{\catcode `\\12\catcode `\$12\catcode
  `\&12\catcode `\#12\catcode `\^12\catcode `\_12\catcode `\%12\relax}%
\providecommand \@@startlink[1]{}%
\providecommand \@@endlink[0]{}%
\providecommand \url  [0]{\begingroup\@sanitize@url \@url }%
\providecommand \@url [1]{\endgroup\@href {#1}{\urlprefix }}%
\providecommand \urlprefix  [0]{URL }%
\providecommand \Eprint [0]{\href }%
\providecommand \doibase [0]{http://dx.doi.org/}%
\providecommand \selectlanguage [0]{\@gobble}%
\providecommand \bibinfo  [0]{\@secondoftwo}%
\providecommand \bibfield  [0]{\@secondoftwo}%
\providecommand \translation [1]{[#1]}%
\providecommand \BibitemOpen [0]{}%
\providecommand \bibitemStop [0]{}%
\providecommand \bibitemNoStop [0]{.\EOS\space}%
\providecommand \EOS [0]{\spacefactor3000\relax}%
\providecommand \BibitemShut  [1]{\csname bibitem#1\endcsname}%
\let\auto@bib@innerbib\@empty
%</preamble>
\bibitem [{\citenamefont {Misner}\ \emph {et~al.}(1973)\citenamefont {Misner},
  \citenamefont {Thorne}, \citenamefont {Wheeler} \emph
  {et~al.}}]{misner1973gravitation}%
  \BibitemOpen
  \bibfield  {author} {\bibinfo {author} {\bibfnamefont {C.~W.}\ \bibnamefont
  {Misner}}, \bibinfo {author} {\bibfnamefont {K.~S.}\ \bibnamefont {Thorne}},
  \bibinfo {author} {\bibfnamefont {J.~A.}\ \bibnamefont {Wheeler}},  \emph
  {et~al.},\ }\href@noop {} {\emph {\bibinfo {title} {Gravitation}}}\ (\bibinfo
   {publisher} {Macmillan},\ \bibinfo {year} {1973})\BibitemShut {NoStop}%
\bibitem [{\citenamefont {Hawking}(1974)}]{hawking1974black}%
  \BibitemOpen
  \bibfield  {author} {\bibinfo {author} {\bibfnamefont {S.~W.}\ \bibnamefont
  {Hawking}},\ }\href@noop {} {\bibfield  {journal} {\bibinfo  {journal}
  {Nature}\ }\textbf {\bibinfo {volume} {248}},\ \bibinfo {pages} {30}
  (\bibinfo {year} {1974})}\BibitemShut {NoStop}%
\bibitem [{\citenamefont
  {Schwarzschild}(1916)}]{schwarzschild1916gravitationsfeld}%
  \BibitemOpen
  \bibfield  {author} {\bibinfo {author} {\bibfnamefont {K.}~\bibnamefont
  {Schwarzschild}},\ }\href@noop {} {\bibfield  {journal} {\bibinfo  {journal}
  {Sitzungsberichte der K\"oniglich Preu\ss ischen Akademie der
  Wissenschaften}\ ,\ \bibinfo {pages} {424}} (\bibinfo {year}
  {1916})}\BibitemShut {NoStop}%
\bibitem [{\citenamefont {Reissner}(1916)}]{reissner1916eigengravitation}%
  \BibitemOpen
  \bibfield  {author} {\bibinfo {author} {\bibfnamefont {H.}~\bibnamefont
  {Reissner}},\ }\href@noop {} {\bibfield  {journal} {\bibinfo  {journal}
  {Annalen der Physik}\ }\textbf {\bibinfo {volume} {355}},\ \bibinfo {pages}
  {106} (\bibinfo {year} {1916})}\BibitemShut {NoStop}%
\bibitem [{\citenamefont {Weyl}(1917)}]{weyl1917gravitationstheorie}%
  \BibitemOpen
  \bibfield  {author} {\bibinfo {author} {\bibfnamefont {H.}~\bibnamefont
  {Weyl}},\ }\href@noop {} {\bibfield  {journal} {\bibinfo  {journal} {Annalen
  der Physik}\ }\textbf {\bibinfo {volume} {359}},\ \bibinfo {pages} {117}
  (\bibinfo {year} {1917})}\BibitemShut {NoStop}%
\bibitem [{\citenamefont {Nordstr{\"o}m}(1918)}]{nordstrom1918energy}%
  \BibitemOpen
  \bibfield  {author} {\bibinfo {author} {\bibfnamefont {G.}~\bibnamefont
  {Nordstr{\"o}m}},\ }\href@noop {} {\bibfield  {journal} {\bibinfo  {journal}
  {Bulletin van de Koninklijke Nederlandse Academie van Wetenschappen}\
  }\textbf {\bibinfo {volume} {20}},\ \bibinfo {pages} {1238} (\bibinfo {year}
  {1918})}\BibitemShut {NoStop}%
\bibitem [{\citenamefont {Jeffery}(1921)}]{jeffery1921field}%
  \BibitemOpen
  \bibfield  {author} {\bibinfo {author} {\bibfnamefont {G.~B.}\ \bibnamefont
  {Jeffery}},\ }\href@noop {} {\bibfield  {journal} {\bibinfo  {journal}
  {Proceedings of the Royal Society of London. Series A}\ }\textbf {\bibinfo
  {volume} {99}},\ \bibinfo {pages} {123} (\bibinfo {year} {1921})}\BibitemShut
  {NoStop}%
\bibitem [{\citenamefont {Newman}\ and\ \citenamefont
  {Janis}(1965)}]{newman1965note}%
  \BibitemOpen
  \bibfield  {author} {\bibinfo {author} {\bibfnamefont {E.~T.}\ \bibnamefont
  {Newman}}\ and\ \bibinfo {author} {\bibfnamefont {A.}~\bibnamefont {Janis}},\
  }\href@noop {} {\bibfield  {journal} {\bibinfo  {journal} {Journal of
  Mathematical Physics}\ }\textbf {\bibinfo {volume} {6}},\ \bibinfo {pages}
  {915} (\bibinfo {year} {1965})}\BibitemShut {NoStop}%
\bibitem [{\citenamefont {Kerr}(1963)}]{kerr1963gravitational}%
  \BibitemOpen
  \bibfield  {author} {\bibinfo {author} {\bibfnamefont {R.~P.}\ \bibnamefont
  {Kerr}},\ }\href@noop {} {\bibfield  {journal} {\bibinfo  {journal} {Physical
  Review Letters}\ }\textbf {\bibinfo {volume} {11}},\ \bibinfo {pages} {237}
  (\bibinfo {year} {1963})}\BibitemShut {NoStop}%
\bibitem [{\citenamefont {Nobel-Prize-Committee}(2020)}]{nobel2020}%
  \BibitemOpen
  \bibfield  {author} {\bibinfo {author} {\bibnamefont
  {Nobel-Prize-Committee}},\ }\href@noop {} {\bibfield  {journal} {\bibinfo
  {journal} {https://www.nobelprize.org/\\prizes/physics/2020/summary/}\ }
  (\bibinfo {year} {2020})}\BibitemShut {NoStop}%
\bibitem [{\citenamefont {Rovelli}(2021)}]{rovelliEPnews2021}%
  \BibitemOpen
  \bibfield  {author} {\bibinfo {author} {\bibfnamefont {C.}~\bibnamefont
  {Rovelli}},\ }\href@noop {} {\bibfield  {journal} {\bibinfo  {journal}
  {Europhysics News}\ }\textbf {\bibinfo {volume} {52}},\ \bibinfo {pages} {16}
  (\bibinfo {year} {2021})}\BibitemShut {NoStop}%
\bibitem [{\citenamefont {Bardeen}(1968)}]{bardeen1968non}%
  \BibitemOpen
  \bibfield  {author} {\bibinfo {author} {\bibfnamefont {J.~M.}\ \bibnamefont
  {Bardeen}},\ }in\ \href@noop {} {\emph {\bibinfo {booktitle} {Proc. Int.
  Conf. GR5, Tbilisi}}},\ Vol.\ \bibinfo {volume} {174}\ (\bibinfo {year}
  {1968})\BibitemShut {NoStop}%
\bibitem [{\citenamefont {Hayward}(2006)}]{hayward2006formation}%
  \BibitemOpen
  \bibfield  {author} {\bibinfo {author} {\bibfnamefont {S.~A.}\ \bibnamefont
  {Hayward}},\ }\href@noop {} {\bibfield  {journal} {\bibinfo  {journal}
  {Physical review letters}\ }\textbf {\bibinfo {volume} {96}},\ \bibinfo
  {pages} {031103} (\bibinfo {year} {2006})}\BibitemShut {NoStop}%
\bibitem [{\citenamefont {Nieuwenhuizen}(2008)}]{nieuwenhuizen2008exact}%
  \BibitemOpen
  \bibfield  {author} {\bibinfo {author} {\bibfnamefont {T.~M.}\ \bibnamefont
  {Nieuwenhuizen}},\ }\href@noop {} {\bibfield  {journal} {\bibinfo  {journal}
  {Fluctuation and Noise Letters}\ }\textbf {\bibinfo {volume} {8}},\ \bibinfo
  {pages} {L141} (\bibinfo {year} {2008})}\BibitemShut {NoStop}%
\bibitem [{\citenamefont {Nieuwenhuizen}\ and\ \citenamefont
  {{\v{S}}pi{\v{c}}ka}(2010)}]{nieuwenhuizen2010bose}%
  \BibitemOpen
  \bibfield  {author} {\bibinfo {author} {\bibfnamefont {T.~M.}\ \bibnamefont
  {Nieuwenhuizen}}\ and\ \bibinfo {author} {\bibfnamefont {V.}~\bibnamefont
  {{\v{S}}pi{\v{c}}ka}},\ }\href@noop {} {\bibfield  {journal} {\bibinfo
  {journal} {Physica E: Low-dimensional Systems and Nanostructures}\ }\textbf
  {\bibinfo {volume} {42}},\ \bibinfo {pages} {256} (\bibinfo {year}
  {2010})}\BibitemShut {NoStop}%
\bibitem [{\citenamefont {Frolov}(2014)}]{frolov2014information}%
  \BibitemOpen
  \bibfield  {author} {\bibinfo {author} {\bibfnamefont {V.~P.}\ \bibnamefont
  {Frolov}},\ }\href@noop {} {\bibfield  {journal} {\bibinfo  {journal}
  {Journal of High Energy Physics}\ }\textbf {\bibinfo {volume} {2014}},\
  \bibinfo {pages} {49} (\bibinfo {year} {2014})}\BibitemShut {NoStop}%
\bibitem [{\citenamefont {Casadio}\ \emph {et~al.}(2015)\citenamefont
  {Casadio}, \citenamefont {Giugno}, \citenamefont {Micu},\ and\ \citenamefont
  {Orlandi}}]{casadio2015thermal}%
  \BibitemOpen
  \bibfield  {author} {\bibinfo {author} {\bibfnamefont {R.}~\bibnamefont
  {Casadio}}, \bibinfo {author} {\bibfnamefont {A.}~\bibnamefont {Giugno}},
  \bibinfo {author} {\bibfnamefont {O.}~\bibnamefont {Micu}}, \ and\ \bibinfo
  {author} {\bibfnamefont {A.}~\bibnamefont {Orlandi}},\ }\href@noop {}
  {\bibfield  {journal} {\bibinfo  {journal} {Entropy}\ }\textbf {\bibinfo
  {volume} {17}},\ \bibinfo {pages} {6893} (\bibinfo {year}
  {2015})}\BibitemShut {NoStop}%
\bibitem [{\citenamefont {Mazur}\ and\ \citenamefont
  {Mottola}(2015)}]{mazur2015surface}%
  \BibitemOpen
  \bibfield  {author} {\bibinfo {author} {\bibfnamefont {P.~O.}\ \bibnamefont
  {Mazur}}\ and\ \bibinfo {author} {\bibfnamefont {E.}~\bibnamefont
  {Mottola}},\ }\href@noop {} {\bibfield  {journal} {\bibinfo  {journal}
  {Classical and Quantum Gravity}\ }\textbf {\bibinfo {volume} {32}},\ \bibinfo
  {pages} {215024} (\bibinfo {year} {2015})}\BibitemShut {NoStop}%
\bibitem [{\citenamefont {Simpson}\ and\ \citenamefont
  {Visser}(2020)}]{simpson2020regular}%
  \BibitemOpen
  \bibfield  {author} {\bibinfo {author} {\bibfnamefont {A.}~\bibnamefont
  {Simpson}}\ and\ \bibinfo {author} {\bibfnamefont {M.}~\bibnamefont
  {Visser}},\ }\href@noop {} {\bibfield  {journal} {\bibinfo  {journal}
  {Universe}\ }\textbf {\bibinfo {volume} {6}},\ \bibinfo {pages} {8} (\bibinfo
  {year} {2020})}\BibitemShut {NoStop}%
\bibitem [{\citenamefont {Ay{\'o}n-Beato}\ and\ \citenamefont
  {Garc{\'i}a}(2000)}]{ayon2000bardeen}%
  \BibitemOpen
  \bibfield  {author} {\bibinfo {author} {\bibfnamefont {E.}~\bibnamefont
  {Ay{\'o}n-Beato}}\ and\ \bibinfo {author} {\bibfnamefont {A.}~\bibnamefont
  {Garc{\'i}a}},\ }\href@noop {} {\bibfield  {journal} {\bibinfo  {journal}
  {Physics Letters B}\ }\textbf {\bibinfo {volume} {493}},\ \bibinfo {pages}
  {149} (\bibinfo {year} {2000})}\BibitemShut {NoStop}%
\bibitem [{\citenamefont {CMS-Collaboration}(2021)}]{CMSPreliminary}%
  \BibitemOpen
  \bibfield  {author} {\bibinfo {author} {\bibnamefont {CMS-Collaboration}},\
  }\href@noop {} {\bibfield  {journal} {\bibinfo  {journal}
  {https://twiki.cern.ch/twiki/bin\\/view/CMSPublic/PhysicsResultsCombined}\ }
  (\bibinfo {year} {2021})}\BibitemShut {NoStop}%
\bibitem [{\citenamefont {Carballo-Rubio}\ \emph {et~al.}(2020)\citenamefont
  {Carballo-Rubio}, \citenamefont {Di~Filippo}, \citenamefont {Liberati},\ and\
  \citenamefont {Visser}}]{carballo2020opening}%
  \BibitemOpen
  \bibfield  {author} {\bibinfo {author} {\bibfnamefont {R.}~\bibnamefont
  {Carballo-Rubio}}, \bibinfo {author} {\bibfnamefont {F.}~\bibnamefont
  {Di~Filippo}}, \bibinfo {author} {\bibfnamefont {S.}~\bibnamefont
  {Liberati}}, \ and\ \bibinfo {author} {\bibfnamefont {M.}~\bibnamefont
  {Visser}},\ }\href@noop {} {\bibfield  {journal} {\bibinfo  {journal}
  {Classical and Quantum Gravity}\ }\textbf {\bibinfo {volume} {37}},\ \bibinfo
  {pages} {145005} (\bibinfo {year} {2020})}\BibitemShut {NoStop}%
\bibitem [{\citenamefont {Painlev{\'e}}(1921)}]{painleve1921mecanique}%
  \BibitemOpen
  \bibfield  {author} {\bibinfo {author} {\bibfnamefont {P.}~\bibnamefont
  {Painlev{\'e}}},\ }\href@noop {} {\bibfield  {journal} {\bibinfo  {journal}
  {CR}\ }\textbf {\bibinfo {volume} {173}},\ \bibinfo {pages} {677} (\bibinfo
  {year} {1921})}\BibitemShut {NoStop}%
\bibitem [{\citenamefont {Gullstrand}(1922)}]{gullstrand1922allgemeine}%
  \BibitemOpen
  \bibfield  {author} {\bibinfo {author} {\bibfnamefont {A.}~\bibnamefont
  {Gullstrand}},\ }\href@noop {} {\emph {\bibinfo {title} {Allgemeine
  L{\"o}sung des statischen Eink{\"o}rperproblems in der Einsteinschen
  Gravitationstheorie}}}\ (\bibinfo  {publisher} {Almqvist \& Wiksell},\
  \bibinfo {year} {1922})\BibitemShut {NoStop}%
\bibitem [{\citenamefont {Landau}\ and\ \citenamefont
  {Lifshitz}(1975)}]{landau1975classical}%
  \BibitemOpen
  \bibfield  {author} {\bibinfo {author} {\bibfnamefont {L.~D.}\ \bibnamefont
  {Landau}}\ and\ \bibinfo {author} {\bibfnamefont {E.~M.}\ \bibnamefont
  {Lifshitz}},\ }\href@noop {} {\bibfield  {journal} {\bibinfo  {journal}
  {Course of Theoretical Physics}\ }\textbf {\bibinfo {volume} {2}} (\bibinfo
  {year} {1975})}\BibitemShut {NoStop}%
\bibitem [{\citenamefont {Babak}\ and\ \citenamefont
  {Grishchuk}(1999)}]{babak1999energy}%
  \BibitemOpen
  \bibfield  {author} {\bibinfo {author} {\bibfnamefont {S.~V.}\ \bibnamefont
  {Babak}}\ and\ \bibinfo {author} {\bibfnamefont {L.~P.}\ \bibnamefont
  {Grishchuk}},\ }\href@noop {} {\bibfield  {journal} {\bibinfo  {journal}
  {Physical Review D}\ }\textbf {\bibinfo {volume} {61}},\ \bibinfo {pages}
  {024038} (\bibinfo {year} {1999})}\BibitemShut {NoStop}%
\bibitem [{\citenamefont {Nieuwenhuizen}(2007)}]{nieuwenhuizen2007einstein}%
  \BibitemOpen
  \bibfield  {author} {\bibinfo {author} {\bibfnamefont {T.~M.}\ \bibnamefont
  {Nieuwenhuizen}},\ }\href@noop {} {\bibfield  {journal} {\bibinfo  {journal}
  {Europhysics Letters}\ }\textbf {\bibinfo {volume} {78}},\ \bibinfo {pages}
  {10010} (\bibinfo {year} {2007})}\BibitemShut {NoStop}%
\bibitem [{\citenamefont {Pitaevskii}\ and\ \citenamefont
  {Stringari}(2016)}]{pitaevskii2016bose}%
  \BibitemOpen
  \bibfield  {author} {\bibinfo {author} {\bibfnamefont {L.}~\bibnamefont
  {Pitaevskii}}\ and\ \bibinfo {author} {\bibfnamefont {S.}~\bibnamefont
  {Stringari}},\ }\href@noop {} {\emph {\bibinfo {title} {Bose-Einstein
  condensation and superfluidity}}},\ Vol.\ \bibinfo {volume} {164}\ (\bibinfo
  {publisher} {Oxford University Press},\ \bibinfo {year} {2016})\BibitemShut
  {NoStop}%
\bibitem [{\citenamefont {Peskin}(2018)}]{peskin2018introduction}%
  \BibitemOpen
  \bibfield  {author} {\bibinfo {author} {\bibfnamefont {M.~E.}\ \bibnamefont
  {Peskin}},\ }\href@noop {} {\emph {\bibinfo {title} {An introduction to
  quantum field theory}}}\ (\bibinfo  {publisher} {CRC press},\ \bibinfo {year}
  {2018})\BibitemShut {NoStop}%
\bibitem [{\citenamefont {Ghosh}\ \emph {et~al.}(2021)\citenamefont {Ghosh},
  \citenamefont {Thalapillil},\ and\ \citenamefont
  {Ullah}}]{ghosh2021astrophysical}%
  \BibitemOpen
  \bibfield  {author} {\bibinfo {author} {\bibfnamefont {D.}~\bibnamefont
  {Ghosh}}, \bibinfo {author} {\bibfnamefont {A.}~\bibnamefont {Thalapillil}},
  \ and\ \bibinfo {author} {\bibfnamefont {F.}~\bibnamefont {Ullah}},\
  }\href@noop {} {\bibfield  {journal} {\bibinfo  {journal} {Physical Review
  D}\ }\textbf {\bibinfo {volume} {103}},\ \bibinfo {pages} {023006} (\bibinfo
  {year} {2021})}\BibitemShut {NoStop}%
\bibitem [{\citenamefont {GRB-Coordinates-Network}(019a)}]{Ligo2019a}%
  \BibitemOpen
  \bibfield  {author} {\bibinfo {author} {\bibnamefont
  {GRB-Coordinates-Network}},\ }\href@noop {} {\bibfield  {journal} {\bibinfo
  {journal} {Circular Service}\ }\textbf {\bibinfo {volume} {24168}} (\bibinfo
  {year} {2019a})}\BibitemShut {NoStop}%
\bibitem [{\citenamefont {GRB-Coordinates-Network}(019b)}]{Ligo2019b}%
  \BibitemOpen
  \bibfield  {author} {\bibinfo {author} {\bibnamefont
  {GRB-Coordinates-Network}},\ }\href@noop {} {\bibfield  {journal} {\bibinfo
  {journal} {Circular Service}\ }\textbf {\bibinfo {volume} {24411}} (\bibinfo
  {year} {2019b})}\BibitemShut {NoStop}%
\bibitem [{\citenamefont {Lattimer}(2019)}]{lattimer2019properties}%
  \BibitemOpen
  \bibfield  {author} {\bibinfo {author} {\bibfnamefont {J.~M.}\ \bibnamefont
  {Lattimer}},\ }\href@noop {} {\bibfield  {journal} {\bibinfo  {journal}
  {arXiv preprint arXiv:1908.03622}\ } (\bibinfo {year} {2019})}\BibitemShut
  {NoStop}%
\end{thebibliography}
\end{document}